\documentclass[
    ,final            
  ]
  {aipproc}

\layoutstyle{6x9}

\begin{document}

\title{New Black Widows and Redbacks in the Galactic Field}


\author{Mallory S.E. Roberts}{
  address={Eureka Scientific, Inc. Oakland, CA USA}
  ,altaddress={with the Fermi Pulsar Search and GBT Drift Scan Survey Collaborations} 
}



\begin{abstract}
There has recently been a large increase in the number of known eclipsing radio millisecond pulsars 
in the Galactic field, many of which are associated with $Fermi$ $\gamma$-ray sources. All 
are in tight binaries  ($P_b < 24$~hr) many of which are classical ``black widows" with very 
low mass companions ($M_c << 0.1 M_{\odot}$) but some are ``redbacks" with probably non-degenerate 
low mass companions ($M_c \sim 0.2 M_{\odot}$). I review the new discoveries, briefly 
discuss the distance uncertainties and the implications for high-energy emission.
\end{abstract}

\maketitle


\section{Black Widows and Redbacks}

There is a subclass of MSPs in tight orbits ($P_b < 24$~hr) which have very low mass
companions ($M_c < 0.05 M_{\odot}$) and tend to exhibit eclipses. The first of these discovered
was PSR B1957+20, a 1.6~ms pulsar in a 9.2~hr orbit around a $\sim 0.02 M_{\odot}$ companion \cite{fst88}.  The radio pulsar regularly eclipses over $\sim 10$\% of the orbit, with the pulses
becoming highly scattered at eclipse ingress and egress, which is direct evidence for significant
amounts of intrabinary material \cite{fbb+90}. The optical lightcurve of the companion
shows large orbital variation, with a peak visual magnitude of $\sim 20.3$. This demonstrates that
the pulsar is heating the companion. $H\alpha$ imaging revealed a nebular bow shock, which is
direct evidence for a strong pulsar wind \cite{kh88}. $Chandra$ X-ray imaging shows a point source and a nebular tail pointing in the opposite direction of the pulsar motion \cite{sgk+03}. 
The X-rays from the point source are orbitally modulated and have a power-law spectrum \cite{hb07}.
The interpretation of these data is that either particle or $\gamma$ radiation from the pulsar is
ablating the companion, and will perhaps eventually evaporate it leaving an isolated pulsar. The
compactness of the orbit and the relatively high spin down energy of the pulsar ($\dot E \sim 10^{35}$erg/s) gives
rise to this phenomenum.
Because the pulsar seems to be destroying its companion, it is called the Black Widow pulsar.

Between 1988 and 2007, only two other eclipsing MSPs with very low mass companions were discovered
in the Galactic field; J2051$-$0827 \cite{sbl+96} and J0610$-$2100 \cite{bjd+06}. Both have significantly lower $\dot E/d^2$
than B1957+20, assuming their respective NE2001 distances \cite{cl02}. 
Many more ``black widows" have been found in globular clusters. In addition, a related class of
eclipsing MSPs with short orbital periods have been found, the first by
Parkes in NGC 6397 \cite{dpm+01}. These have companion masses of a
few tenths of a solar mass. In cases where an optical counterpart has been identified, they often
appear to have non-degenerate companions, suggesting they are still in the late stages of recycling.
I will therefore refer to these systems as ``redbacks", the Australian cousin to the North American
black widow spider. A total of 18 black widows and 12 redbacks are listed on P. Freire's website 
of globular cluster pulsars ({\tt  http://www.naic.edu/$\sim$pfreire/GCpsr.html}). However, globular clusters are distant objects, and measurements of
$\dot P$ are strongly affected by acceleration in the cluster gravitational potential. Therefore, 
nearby eclipsing MSPs in the Galactic field are highly desirable for studies of 
relativistic shock emission in these systems.

\section{High Energy Emission from an Intrabinary Shock}

The pulsar wind can interact with ablated material producing an intrabinary shock front whose
orientation changes in respect to Earth as a function of orbital phase. It is this shock front
which obscures the pulsed emission during the eclipses.
Arons and Tavani \cite{at93} 
developed a model of high energy shock emission for the original Black Widow. They predict that electrons 
could be accelerated to energies as high as 3 TeV in this system. Although the
orbit is circular, the high energy emission could be orbitally modulated due to obscuration by the
shock, intrinsic beaming of the particle acceleration and emission by the magnetic field orientation
in the shock, and from Doppler boosting. The shock distance of only a few light seconds from
the pulsar implies that the B field and magnetization parameter $\sigma$ may be relatively high
compared to the termination shock of an ordinary pulsar wind nebula. While the shock luminosity
can depend on the shock distance, fraction of wind involved, pulsar magnetic field, optical emission
of the companion, magnetization of the wind, and the ion fraction of the wind, the most
important factor in determining the total emission is still simply $\dot E$.


An important question in estimating expected high-energy fluxes is how reliable
are NE2001 distance estimates? Accurate parallax measurements (i.e ones with high significance 
measurements that are minimally affected by the Lutz-Kelker bias \cite{vlm10}) since 2001 show that the 20\% error
estimate is generally pretty good in the Galactic plane, but at mid to high Galactic latitudes
where MSPs tend to be found, the NE2001 model systematically underestimates the distance (Fig. 1),
often by as much as a factor of 2 \cite{cbv+09}. This has been attributed to a poor 
estimate of the Galactic scale height for the gas in the model. This would imply estimates of $\dot E/d^2$
for new eclipsing systems in the field are likely overestimated by a factor of $\sim 2-4$. However, 
we now know that MSP masses can be as high as $2 M_{\odot}$, and the radii are likely larger than 
10~km \cite{dpr+10} and so the moment of inertia may be about a factor of 2 larger than 
the canonical value used in calculating $\dot E$, countering the systematic effect of the distance underestimate.  
\begin{figure}[h!]
  \includegraphics[height=.33\textheight]{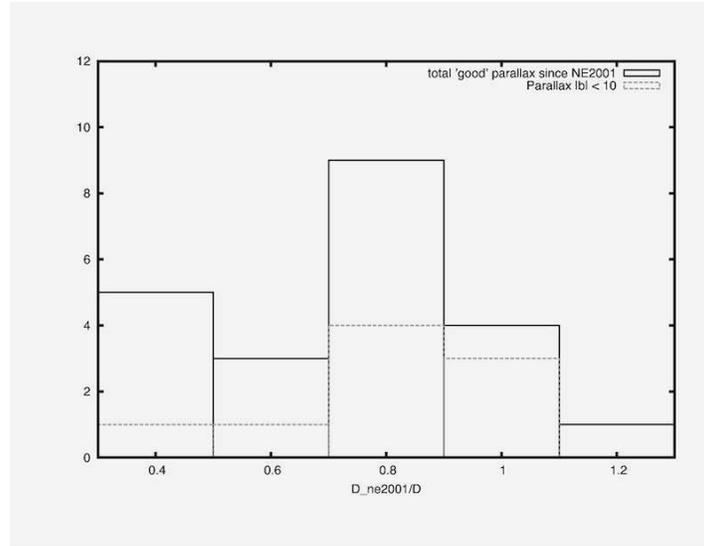}
  \caption{Histogram of the ratios of NE2001 distances to high confidence parallax measurements of pulsars
since 2001 \cite{vlm10}. The solid black histogram is all measurements, the dashed grey is for pulsars at Galactic latitude 
$|b| < 10^{\circ}$}
\end{figure}

\section{New Discoveries in the Galactic Field}

New radio surveys designed to be sensitive to very fast pulsars in tight binaries have recently
greatly increased the number of known MSPs in the Galactic field, including the number of black
widows and redbacks. Large scale sky surveys with GBT and Parkes have so far discovered 2 new black widows and
2 new redbacks. The best studied of these is J1023+0038, discovered in the GBT drift scan survey
\cite{asr+09,a+11}.  This 1.69~ms eclipsing pulsar in a 4.8~hr orbit around
a $\sim 0.2M_{\odot}$ companion is especially interesting because optical observations of the companion
star in 2001 showed spectral evidence for an accretion disk. This further justifies the name ``redback" for these systems, since this spider is one of only two known species where the male actively aids the female
in eating him while mating.

The most productive means of finding field MSPs is proving to be searches of $Fermi$ $\gamma$-ray
sources. As of this time roughly 30 new MSPs have been discovered by searches with the Parkes,
Nancay, Effelsberg,  and especially the Green Bank Telescopes over the last 2 years \cite{rrc+10,b+11,h+11,blm+11}, including at least 4 black widows and 2 redbacks (the number is
changing on an almost daily basis). Most show clear eclipses, including PSR J1810+17 with a spin period of 1.66~ms, second
only to B1957+20 in spin period, but with a much shorter orbit of 3.6~hr, and PSR J2215+51 with 
a spin period of 2.61~ms and a minimum companion mass of $\sim 0.22 M_{\odot}$. The two eclipsing 
sytems from the drift scan survey are also Fermi sources, as is the original black widow.
It should be noted that there is no strong evidence of unpulsed GeV emission in the
Fermi data from these sources yet, with clear $\gamma$-ray pulsations observed from at least three
of the black widows. 

The table lists all the new systems that were announced by the time of the meeting. Many of these
are actually quite bright in radio, and so may be amenable to VLBI parallax measurements.  Most 
are fairly faint Fermi sources, and so disentangling any shock emission in the GeV region from 
pulsed emission will be challenging. However, they may be good candidates for TeV telescopes, 
especially redbacks with non-degenerate companions which can provide a dense field of 
seed photons for inverse Compton scattering. 

This research was partially funded through the Fermi GI program, NASA
grant \#NNG10PB13P








\begin{table}[h!]
\begin{tabular}{lcccccc}
\hline
Pulsar\tablenote{an F indicates a Fermi source} &  $P_s$ & $\dot E/10^{34}$\tablenote{assuming $1.4M_{\odot}$ and 10km radius}  & $d_{NE2001}$ &
$P_B$ & $M_c$\tablenote{assuming $1.4M_{\odot}$ pulsar and $i=90^{\circ}$} & ref. \\ 
 & (ms) & (erg/s) & (kpc) & (hr) & (solar) \\ \hline
\multicolumn{6}{c}{Old Black Widows} \\
\hline
B1957+20 F &  1.61  &  11  & 2.5 & 9.2 & 0.021 & \cite{fbb+90}\\
J0610$-$2100 &  3.86 & 0.23 & 3.5 & 6.9 & 0.025 & \cite{bjd+06} \\ 
J2051$-$0827 & 4.51 & 0.53 & 1.0 & 2.4 & 0.027 & \cite{sbl+96}\\
\hline
\multicolumn{6}{c}{New Black Widows} \\
\hline
J2241$-$52 F & 2.19 & 3.3 & 0.5 & 3.4 & 0.012 & \cite{k+10} \\
J2214+3000 F & 3.12 & 1.9 & 3.6 & 10.0 & 0.014 & \cite{rrc+10} \\
J1745+10 F & 2.65 & 1.3 & 1.3 & 17.5 & 0.014 & \cite{fc11} \\
J0023+09 F & 3.05 & ?? & 0.7 & 3.3 & 0.016 & \cite{h+11} \\
J2256$-$1024 F & 2.29 & 5.2 & 0.6 & 5.1 & 0.034 & \cite{blm+11} \\
J1731$-$1847 & 2.3 & ?? & 2.5 & 7.5 & 0.04 & \cite{b+11} \\
J1810+17 F & 1.66 & ?? & 2.0 & 3.6 & 0.044 & \cite{h+11} \\
\hline
\multicolumn{6}{c}{New Redbacks} \\
\hline
J1023+0038 F & 1.69 &$\sim 5$ & 0.6 & 4.8 & 0.2 & \cite{asr+09}\\
J2215+51 F & 2.61 & ?? & 3.0 & 4.2 & 0.22 & \cite{h+11}\\
J1723$-$28 & 1.86 & ?? & 0.75 & 14.8 & 0.24 & \cite{clm+10} \\
\hline
\end{tabular}
\vspace{-0.5cm}
\caption{Black Widows and Redbacks in the Galactic Field}
\label{tab:bws}
\end{table}
\bibliographystyle{aipproc}   


\end{document}